# Where do highly cited scientists stand? A multi-dimensional analysis of positional distribution in the Top 2% ranking

Yuehai Li[1], Zuguang Gu[1,*]

[1]Institute of Cell and Gene Technology, Shenzhen University of Advanced Technology, Shenzhen, China

*Corresponding author

Email address:

ZG: guzuguang@suat-sz.edu.cn

# Abstract

The Top 2% Scientists list is widely used. Yet current applications either only count the number of scientists in different units or criticize the list's validity as a measure of individual impact. We introduce *positional enrichment*, a scale-free framework that compares observed to expected counts across percentile bins on the list, revealing where a unit's scientists tend to sit along the list rather than how many it contributes. Applying positional enrichment to the 2026 single-year list, we show that countries, institutions, and research subfields each fall into a consistent pattern of three groups with Head-, Middle-, and Tail-enriched profiles. Subfield profiles are globally stable, whereas countries and institutions are heterogeneous across subfields. National and institutional differences are therefore compositional. Therefore, the reason of a country or institution sitting near the head of the list is unlikely its scientists performing better within the same subfields, while it holding more of the subfields that sit near the head. Academic age adds an independent dimension, with stronger enrichment toward the head after 25 years and exceptions in fast-moving subfields such as artificial intelligence. Across country, institution, subfield, and academic age levels, a consistent pattern emerges. Head units contain more Head subfields, and these subfields have a higher proportion of senior scientists with more established citation records, indicating that positional distribution of a unit is jointly organized by disciplinary composition and seniority structure.

# Introduction

The Top 2% Scientists list has become one of the most widely used bibliometric resources in research evaluation [1]. In its newest version released on 2026 October, around 250,000 scientists are ranked by a composite citation score (the c-score) that combines six standardized indicators: total citations, Hirsch h-index, co-authorship-adjusted hm-index, and citations to papers as single, single-or-first, and single-first-or-last author [2], and generates two separate ranking lists for career-long and single-year impact. The list provides a single ordered ranking in which scientists from every country and every research field can be compared on a common metric. In recent years, universities, research institutes, and national agencies have increasingly highlighted their positions in the list as evidence of research strength [3], [4]. The list has become a routine reference in institutional rankings and policy documents [5], [6], and is also used to inform hiring and promotion decisions. The list's influence on institutional reputation and resource allocation has made it more than a bibliometric resource. It is now a benchmark that shapes how institutions present themselves and how national systems assess their positions in the global scientific landscape.

The most common way of using the Top 2% list is to report the number of scientists on it. More in-depth research has focused almost exclusively on the ranking itself, asking whether the list is a valid measure of individual impact. Studies have criticized the biases of highly cited researchers lists more broadly, including the prevalence of hyperprolific authors [7], citations from non-peer-reviewed articles [8], the coverage bias of its underlying publisher database [9], the impact of self-citation [10], and the anomalies that arise from publication-type filtering, temporal coverage [11], and retraction thresholds [12]. Nevertheless, what has received far less attention is the internal structure of the ranking. The list is not a homogeneous block of scientists ordered by a single dimension, while it is a stratified space in which different countries, institutions, and research fields occupy systematically different regions. Two countries can contribute similar numbers of scientists to the list yet differ sharply in where those scientists sit, e.g., one concentrated near the top, and the other near the bottom of the list. Understanding this internal structure, where different units sit, is a precondition for interpreting what the list actually measures.

How are the top scientists located along the list, and what structure underlies their placement? To answer this question, we introduce a statistical framework, *positional enrichment*, for characterizing the internal structure of a ranking. Methodologically, the Top 2% list is divided into ten equal-sized percentile bins, and for each unit (i.e., country, institution, or subfield) we compute the log ratio of observed to expected counts in each bin, where the expected count is the number that the unit would contribute if its scientists were uniformly distributed across the ranking. This measure has two properties that make it suited to the question. First, it is defined against a random baseline**.** The enrichment measures how far a unit's positional distribution departs from uniform placement. Second, it is scale-free**.** Both observed and expected counts scale with the unit's size, thus the enrichment ratio is comparable across units ranging from tens of thousands of scientists to a few hundred. Together, these two properties allow to describe where along the ranking a unit's scientists are concentrated, rather than treating the ranking as a simple count of members.

We applied positional enrichment analysis to the Top 2% list across multiple dimensions: countries, institutions, research fields and their subfields, academic ages, and interactions among these dimensions. We found that every unit type falls into three clearly separated positional profiles, which we labelled Head**,** Middle, and Tail**.** Each label indicates where along the list the unit's dominant enrichment lies. By analyzing the positional enrichment profiles, we have three key findings. First, subfield profiles are globally stable. The same subfield occupies the same region of the ranking whether measured at the global level, within a country, or within an institution. Countries and institutions are heterogeneous across subfields, but subfields are homogeneous across countries and institutions. Second, differences across units are compositional. Head countries and Head institutions sit at the head of the list not because their scientists behave differently within a subfield, but because they contain larger shares of Head subfields. Third, academic age adds an independent axis. Below 25 years, scientists are concentrated at the tail of the list, while above 25 years, they shift toward the head. Together, these findings show that the ranking is not a flat ordering of individual influence but a subfield-organized stratification, in which the position of a country, institution, or scientist is shaped primarily by the subfields they contain.

The c-score and its six components have well-documented limitations. The underlying Scopus database systematically under-covers researchers outside Elsevier journals and outside STEM fields [9]; self-citation can inflate individual rankings [10]; hyperprolific authors and publication-type anomalies can distort the tail of the list [7], [11], and retractions are not incorporated into the ranking [12]. These limitations are serious enough that the list should not be used as a stand-alone measure of research quality. Despite these limitations, we do not attempt to correct, reweight, or revalidate these indicators, nor do we argue that the list is a valid measure of individual impact. Our object of study is the list as it is published, because that is the list used in current evaluations. The question we ask is not whether the list is a valid measure of individual impact, but rather its internal structure and what that structure implies for how the list should be read. Positional enrichment is designed to reveal global patterns at various aggregate levels, across countries, institutions, and fields, and at these levels the known limitations are unlikely to overturn the structure we describe. They affect a small fraction of scientists, and sensitivity analyses already show that the core ranking is highly robust to combined filtering for publication type, temporal coverage, and retraction thresholds [11].

# Data and Conventions

We used the of Top 2% list published on 7 October 2026 [13], where citations are updated to the end of 2025. We used only the single-year list, which contains 249,310 scientists drawn from a global population of 11,619,197 scientists, and rankings are only based on total citations they received in the year of 2025. The list provides classification columns for 171 countries, 28,578 institutions, 20 research fields, 174 sub-fields, and first publication years. For each column, a scientist is assigned a single unique class label. Rankings after removing self-citations were used. The original list was used as the data for this study, and no further filtering was applied. Source code for this study is available at https://github.com/jokergoo/top2ranking.

Throughout the paper, we use *ranking* to denote a unit's (a country's, an institution's or a field's) position in the ordered list, with the convention that a smaller integer value corresponds to a higher position, that is, ranking of 1 is the top of the list. In this study, we also report the *normalized ranking*, defined as the integer ranking divided by the total number of scientists in the list, so that a normalized ranking of 0.01 corresponds to the top 1% of the list.

When a unit's position changes relative to a baseline, for example, an institution's position relative to its national baseline, or a subfield's position relative to its field's baseline, we describe the direction in terms of *ranking improvement* and *ranking drop*. A ranking improvement means the unit moves toward the head of the list and a ranking drop means it moves toward the tail. To reduce ambiguity, when describing a ranking difference, we explicitly state its direction and report only the absolute magnitude of the difference.

# Results

## Positional enrichment

The Top 2% ranking is an ordered list, denoted as $L$. $L$ is divided into ten equal-sized percentile bins $Q_1 \sim Q_{10}$, where $Q_1$ contains the top 10% of scientists and $Q_{10}$ contains the bottom 10%. For a unit $u$ (a country, an institute or a research field) with $k_u$ scientists, the observed number in bin $d$ is denoted as $O_{u,d}$ (obvious $k_u = \sum_{d=1}^{10} O_{u,d}$), and the expected number in the bin under a uniform distribution is $E_u = k_u/10$, then the positional enrichment of unit $u$ in bin $d$, denoted as $r_{u,d}$ is defined as:

$$r_{u,d} = \log_2\left(\frac{O_{\mathrm{u},d}+10}{E_u+10}\right). \tag{1}$$

The additive pseudo-count of 10 stabilizes the ratio for bins with small counts, so that units with few scientists per bin do not produce extreme enrichment values driven by only a few individuals. From the definition, a positive $r_{u,d}$ indicates over-representation in $Q_d$, and a negative value indicates under-representation. The vector $r_u = (r_{u,1}, \ldots, r_{u,10})$ is the positional enrichment profile of the unit.

The number of scientists in unit $u$ in bin $d$, denoted as a random variable $X$, follows a binomial distribution $X \sim \mathrm{B}(N/10, p)$ where $N$ is the total number of scientists on $L$ and $p$ is the probability of a scientist from unit $u$, calculated as $p = k_u/N$, then the mean of $X$ denoted as $\mu$ is calculated

as:

$$\mu = N/10 \cdot p = N/10 \cdot k_u/N = k_u/10. \tag{2}$$

And the enrichment including the pseudo-count has the same form as Formula 1:

$$r_{u,d} = \log_2\left(\frac{O_{u,d}+10}{\mu+10}\right) \ . \tag{3}$$

## Countries exhibit three distinct positional enrichment groups

We computed the positional enrichment profile for each country and applied spherical *k*-means clustering with cosine similarity to separate countries and regions[1] into clusters. Spherical *k*-means clustering captures similar enrichment shapes, i.e., where along the ranking the enrichment is concentrated, regardless of its overall magnitude of enrichment. We ran the clustering 100 times with random initializations and took the consensus solution. The consensus clustering stably separates the countries into three visually distinct blocks in the enrichment heatmap (mean concordance to the consensus labels = 100%, Figure 1), which we labelled *Head*, *Middle*, and *Tail* by their enrichment. Countries with fewer than 200 scientists were excluded, leaving 56 countries in the analysis. The complete positional enrichment table for countries is in Supplementary Table 1.

The Head group is enriched in the leading bins and depleted toward the tail, with strong enrichment in $Q_1 \sim Q_4$. The group is dominated by high-income economies and includes most of the traditional Western countries. They are major English-speaking economies, Western and Northern European countries, as well as Israel. Besides these, there are also five Asian economies that are classified as developing but have high income levels within the Asian region. Hong Kong shows the steepest head-enrichment.

The Middle group is negatively enriched on the top of the list and weakly enriched at the other end, but with strong enrichment in $Q_7 \sim Q_9$. It mainly contains countries from Southern and Eastern Europe, as well as the four largest Asian economies: mainland China, India, Japan, and South Korea. Italy and Spain, which are thought of as major European countries, fall behind on the Top 2% list compared to their European peers such as Germany and the Netherlands. This agrees to our previous finding that the citation preferences of Italy and Spain are isolated from the major Head countries [14]. Last, surprisingly, three developed European countries, Belgium, Austria, and France, have positional enrichment profiles that also fall behind those of other developed Head countries, indicating that the influence of their scientists is not aligned with their economic levels.

---

[1] In the remaining part of this paper, "countries and regions" are described only as countries for simplicity.

The Tail group is depleted in the leading bins, with strong enrichment toward $Q_{10}$. Most of the countries in this group are developing countries in the Middle East, South Asia, Latin America, and Africa, and are basically middle-income and low-income. However, there are exceptions: Ireland and Taiwan fall into the Tail group, possibly because their economic structures significantly differ from their scientific research directions.

In general, Figure 1 illustrates that the position of top scientists is strongly associated with the economic levels of their countries. As the level of economic development rises, more investment is put into science, which increases scientific influence and ultimately pushes scientists toward the top of the list. As more scientific investment is put into hot research fields, we will explore how such top research fields shape the position distribution patterns in the next sections.

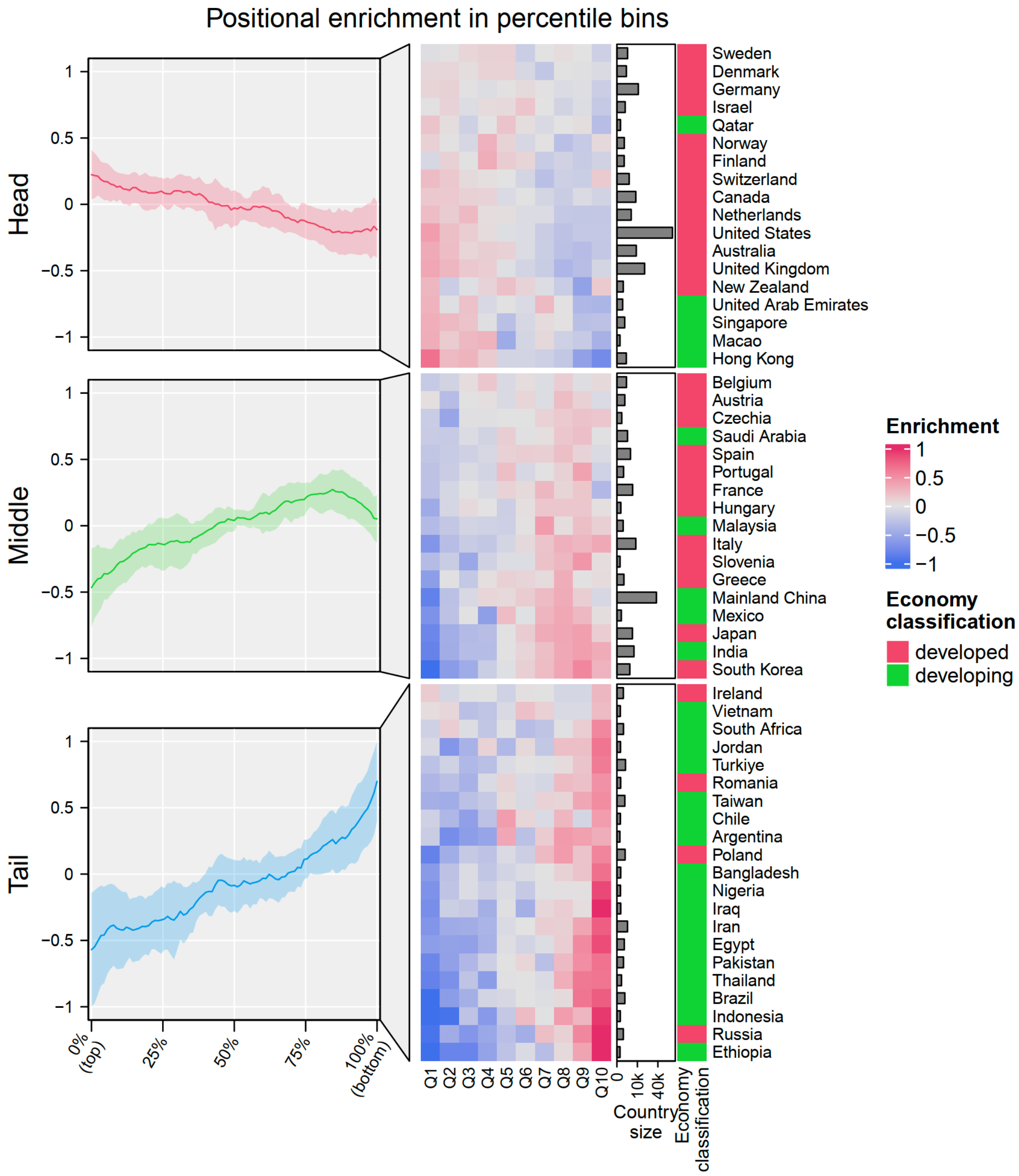

**Figure 1. Positional enrichment of countries.** Line charts illustrate the average enrichment trend within each group. To smooth the distributions, the 10-percentile bins were shifted in steps of 1% along the Top 2% list. Shaded areas around the trend lines represent standard deviations. Barplot of country size is in square root scale. The economy classifications are from the UN Trade and Development data hub[2].

## Institutions inherit the positional enrichment of their countries

Countries are partitioned into institutions. We next ask whether the three-cluster national structure can be reproduced at the level of individual research institutions. We computed the positional enrichment profile with institutions as units, requiring a minimum of 50 scientists per institution, yielding 949 institutions from 45 countries. We then computed the Spearman correlation between an institution's profile and its country's profile. A positive correlation indicates that the institution follows the shape of its country's profile, and a negative value indicates that it deviates in the opposite direction. The positional enrichment table for institutions is in Supplementary Table 2.

[2] https://unctadstat.unctad.org/EN/Classifications.html

Figure 2A splits all institutions by their country's enrichment group. 18 Head-enriched countries contain 578 institutions (60.1% of all institutions), of which 73.4% are positively correlated with their countries. 16 Middle-enriched countries contain 324 institutes (34.1%), of which 89.2% are positively correlated. 11 Tail-enriched countries, although containing only 47 institutions, have 93.6% positive. The Middle and Tail groups are therefore almost uniformly concordant with their national baseline, whereas the Head group contains a substantially larger fraction of institutions whose positional profile runs against the national trend (26.6%). Supplementary Figure 1 illustrates the heatmap of enrichment profiles of all institutions and their countries, which confirms this heterogeneity in Head-enriched countries.

To interpret how countries are composed from their institutions, we also applied consensus spherical *k*-means clustering to the institution profiles. This also yields a stable three-group classification with the same Head-, Middle-, and Tail-enrichment patterns (Supplementary Figure 1). Figure 2B shows that Head-enriched countries are mainly composed of Head-enriched institutions (75.5%), and that as a country's enrichment moves toward the tail of the list, the proportion of Middle- and Tail-enriched institutions increases.

To measure the deviations, we related the institution-country correlation to the mean ranking difference between them (Figure 2C). In Head-enriched countries, lower and eventually negative correlation is associated with institutions whose mean ranking is worse than the national average (i.e., ranking drop). As an institution departs from the national profile, it drifts toward the tail of the list. In Middle- and Tail-enriched countries, the relationship reverses. Lower correlation is associated with institutions whose mean ranking is better than the national average (i.e., ranking improvement).

Although institutions generally show the same directional enrichment as their countries, the reverse enrichment institutions are worth examining. Figure 2D shows institutions in the Middle or Tail countries with a Spearman correlation $< -0.1$ and a normalized ranking improvement $> 0.05$ (which corresponds to an absolute improvement of 12,465 positions). These institutions run opposite to their national profile while ranking substantially better than the national average. They are almost without exception the leading institutions of their national systems, which are Peking University, Tsinghua University, and the Chinese Academy of Sciences, and Westlake University in China, KAUST and KFUPM in Saudi Arabia, Hanyang, Sungkyunkwan, and Sejong in South Korea, the University of Tehran in Iran, and Milano-Bicocca and Vita-Salute San Raffaele in Italy. These institutions represent the most advanced scientific capacity of their countries and they concentrate their scientists at the head of the ranking than their national baseline.

Figure 2E shows the mirror image where institutions in Head countries with a Spearman correlation < -0.55 and a normalized ranking drop > 0.1 (an absolute ranking drop of 24,931 positions)[3]. These institutions fall into three categories, including medical schools and health-science faculties (e.g., University of Wisconsin School of Medicine and Public Health), government and mission-driven laboratories (e.g., Centers for Disease Control and Prevention in the US), and industry and applied-research organizations (e.g., Pfizer Inc.). Their core scientific missions lie outside citation-dense researches. Medical schools publish clinical-related work such as case reports or observational studies that accumulates citations more slowly than basic-research papers [15], [16]. Government and mission-driven laboratories produce reports, standards, and policy-oriented documents that are often cited in non-academic contexts or not indexed in bibliometrics databases [17]. Industry-affiliated scientists face corporate approval processes and intellectual-property constraints before their work can be published. A recent study showed negative correlation between the percentage of industry authors and academic citations [18].

[3] The different cutoffs used for Figure 2D and 2E are mainly for keeping the number of rows in the heatmaps small for readability. The complete data are in Supplementary Table 2.

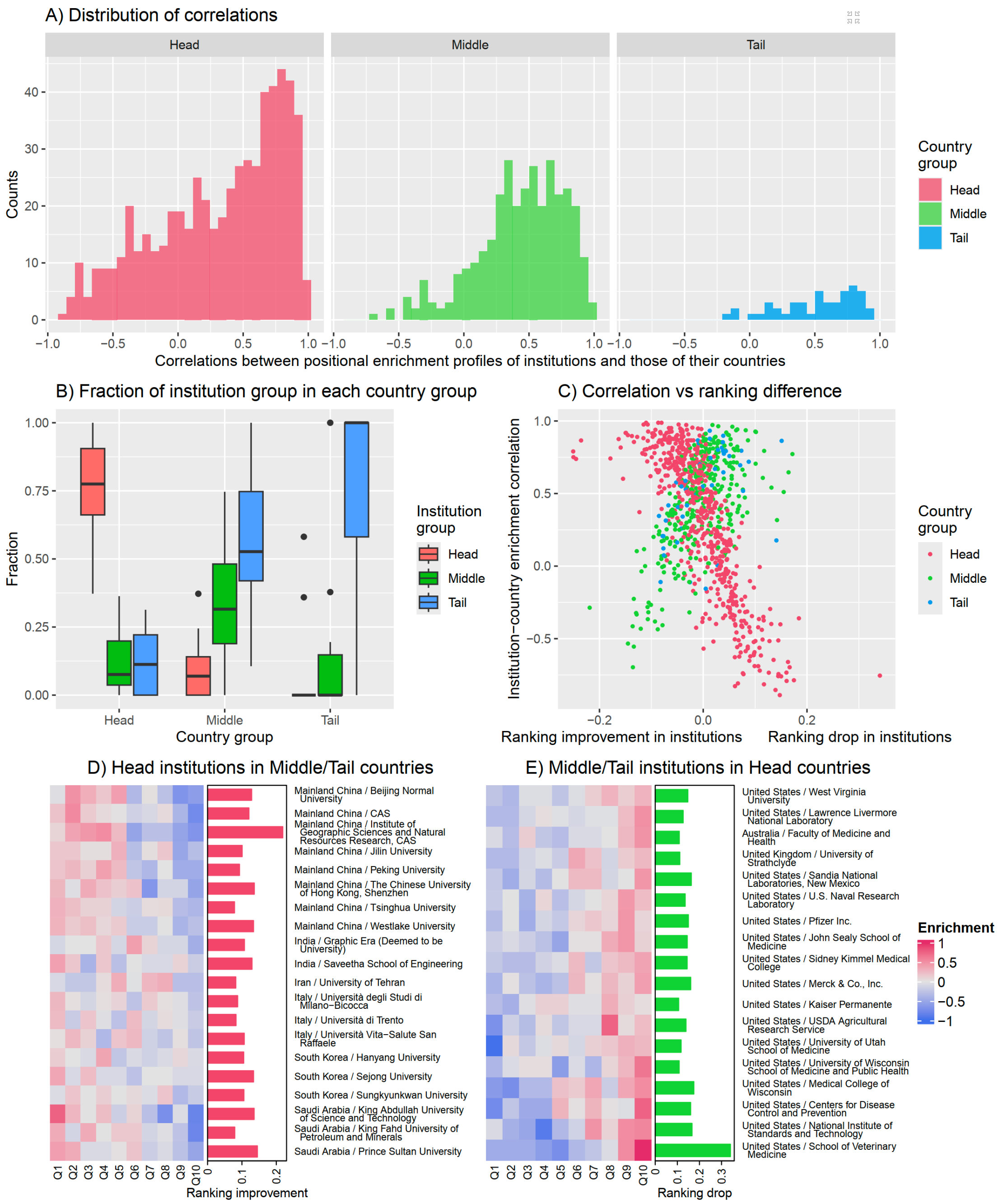

**Figure 2. Positional enrichment of institutions.** (A) Histograms of correlations between positional enrichment profiles of institutions and those of their countries. (B) Fraction of institution groups within each country group. (C) Relationship between institution-country enrichment correlation and the ranking difference between them. Ranking difference is calculated as the average ranking of an institution's scientists minus the average ranking of its country's scientists. (D) Head-enriched institutions in Middle- or Tail-enriched countries. (E) Middle- or Tail-enriched institutions in Head-enriched countries.

## Research fields also split into three enrichment groups, with subfield-level heterogeneity

Using the same method as for countries, we applied the positional enrichment measure to research fields (Figure 3A) and subfields (Figure 3B), and clustered each level into three groups using spherical *k*-means with consensus clustering. The minimum number of scientists was set to 100 for fields and 50 for subfields, yielding 20 fields and 169 subfields. The consensus clustering was almost perfectly stable across 100 clustering runs (100% for fields, 99.8% for subfields).

The field-level enrichment heatmap (Figure 3A) reproduces the three-cluster structure observed from countries, i.e., Head-, Middle-, and Tail-enriched. The subfield-level heatmap (Figure 3B) shows an even clearer three-block structure at finer resolution, indicating that subfields may largely inherit their field's profile. For each field, we asked what fraction of its subfields fall into each profile. The stacked bar plots alongside Figure 3A show that this inheritance is strong. For Head-enriched fields, 67.2% of subfields are themselves Head-enriched (39 out of 58); for Middle- and Tail-enriched fields, 48.4% (15 out of 31) and 57.5% (46 out of 80) of subfields fall into the same profile. Subfield profiles are therefore largely consistent with their parent field, and the three-profile structure holds at the subfield level (Complete tables are in Supplementary Table 3 and 4).

Despite the high consistency of the positional enrichment distributions between fields and their subfields, a certain level of heterogeneity remains (Figure 3A). Figure 3C lists Head-enriched subfields embedded in Middle- or Tail-enriched fields. The list falls into a few coherent groups that share a citation-dense, quantitative, or clinical character. The most typical subfields are quantitative and methodological (e.g., Applied Mathematics and Statistics & Probability within Mathematics & Statistics), followed by physics-related subfields (e.g., Nanoscience & Nanotechnology within Enabling & Strategic Technologies), and a large biomedical and clinical group (e.g., Neurology & Neurosurgery within Clinical Medicine). Nanoscience & Nanotechnology shows the strongest shift to the head of the list (Figure 3C). These subfields accumulate citations faster than the fields in which they are embedded.

Figure 3D lists the opposite cases, Tail-enriched subfields embedded in Head fields, additionally filtered by a normalized ranking drop > 0.1. These are the citation-sparse, qualitative, or organismal parts of the fields. The largest group is humanistic and social-scientific (e.g., Law within Social Sciences, Religions & Theology within Philosophy & Theology, Design Practice & Management within Built Environment & Design), which publish in books and monographs and are evaluated through qualitative rather than citation-based criteria. A second group is organismal and natural-history (e.g., Zoology within Biology, Geology and Oceanography within Earth & Environmental Sciences), which are descriptive and specimen-based, with publication cultures closer to natural history than to the experimental biosciences that dominate their parent fields. A third group is clinical-humanistic and interpretive (e.g., General Psychology & Cognitive Sciences within Psychology & Cognitive Sciences). Zoology shows the steepest ranking decrease. Together, Figures 3C and 3D show that a field is not internally uniform. It may contain both Head-enriched and Tail-enriched subfields, and where a subfield sits depends more on how citation-dense its research field is.

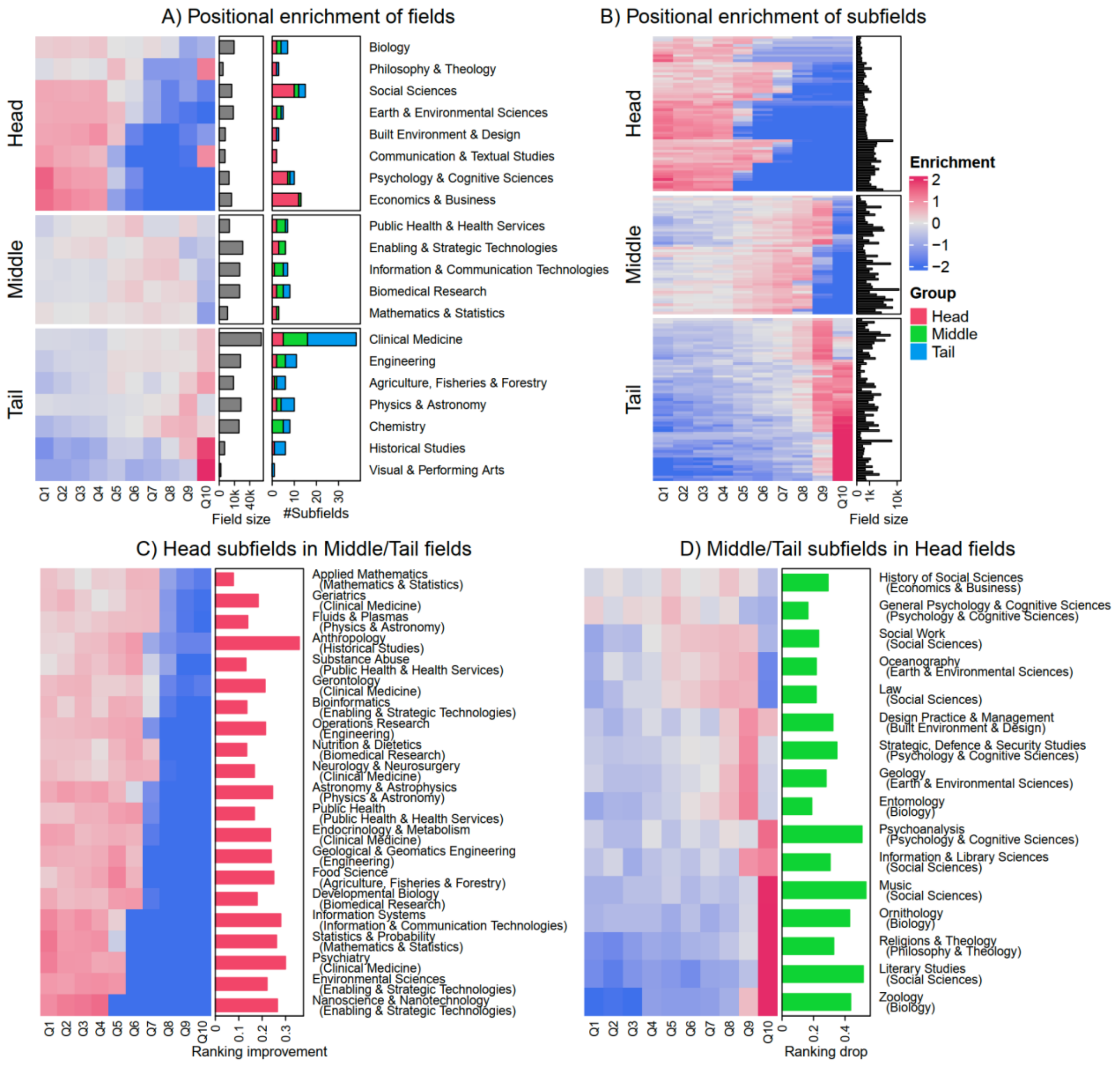

**Figure 3. Positional enrichment of research fields and subfields.** A) Positional enrichment of research fields. B) Positional enrichment of subfields. Scales of barplots in panels A and B are both in square root scale. C) Head-enriched subfields in Middle-/Tail-enriched parent fields. D) Middle-/Tail-enriched subfields in Head parent fields. Text in the parentheses indicates the parent field of each subfield.

## Research subfields organize positional enrichment

We computed the positional enrichment profile of each country × subfield and institution × subfield combination, together with the corresponding global subfield profile, country profile, and institution profile. The minimal number of scientists was set to 50 for country x subfield, yielding 926 combinations from 41 countries and 153 subfields, and to 25 for institution x subfield, yielding 423 combinations from 217 institutions in 22 countries and 56 subfields (Supplementary Table 5 and 6). For both levels, correlations with the subfield profile are almost uniformly positive (Figure 4A and 4B, the “Correlation” column beside the “Subfields” heatmap). Nearly every country × subfield unit (99.7%) and institution × subfield unit (98.6%) follows the positional shape of its subfield, whereas correlations with the country or institution profile are mixed (Figure 4A and 4B, the “Correlation” column beside the “Countries” or “Institutions” heatmap), with a substantial fraction of units deviating in the negative direction (37.8% for country × subfield, and 17.3% for institution × subfield). This indicates that a subfield's positional profile is consistent whether it is measured globally, within a country, or within an institution.

The consistency can be measured quantitatively. For each country, we computed the mean pairwise Spearman correlation of its subfields' enrichment profiles (Figure 4C). The correlations are low (overall mean = 0.125), indicating that a single country's subfields occupy different regions of the ranking. A single country may contain Head, Middle, and Tail subfields simultaneously. Countries with fewer subfields may show higher mean correlations, reflecting their lower internal heterogeneity. For each subfield, we also computed the mean pairwise Spearman correlation of its countries' enrichment profiles (Figure 4D). The correlations are high (overall mean = 0.795), indicating that a subfield's positional profile is highly consistent across countries regardless of their group. Head, Middle, and Tail countries show similar enrichment profiles within the same subfield. The consistency is even higher for Head subfields (overall mean = 0.916) and is stable across subfield sizes. Countries are therefore heterogeneous across subfields, and subfields are homogeneous across countries.

This asymmetry of countries and subfields implies that the separation of Head, Middle, and Tail countries reflects differences in their subfield composition. Figure 4G shows, for each country group, the share of its subfields that fall into the Head, Middle, and Tail groups. Head countries have the highest share of Head subfields (on average 28.8%), Middle countries an intermediate share (18.4%), and Tail countries the lowest (15.8%). The same ordering holds for the other two country groups. Middle countries have the highest share of Middle subfield (47.2%), Tail countries have the highest Tail subfields (43.5%). The national separation is therefore a composition effect, which is driven by which subfields a country contains rather than by how well its scientists rank within a given subfield.

Figures 4E-4F repeat the analysis at the institutional level. For each institution, we computed

the mean pairwise Spearman correlation of its subfields' enrichment profiles (Figure 4E). The correlations are low (overall mean = 0.081) and become more variable as institution size decreases. An institution × subfield unit is more likely to be much smaller than a country × subfield unit, and smaller units are more likely to show low internal heterogeneity, which increases the variability of their correlation estimates across units. For each subfield, we also computed the mean pairwise Spearman correlation of its institutions' enrichment profiles (Figure 4F). The correlations are high (overall mean = 0.634). Although lower than at the country level, especially for Middle (overall mean = 0.496) and Tail subfields (overall mean = 0.627), Head subfields still show strong consistency (overall mean = 0.821).

To further investigate the combinational relation among countries, institutions and subfields, we restricted the analysis to institutions only in United States and mainland China, two large countries from the Head and Middle groups that together contribute the majority of institutions in the analysis (295 in United States and 142 in mainland China, out of 949 institutions with at least 50 scientists). Figure 4H reproduces the pattern from Figure 4G, that is, institutions in each enrichment group contain a larger fraction of subfields in the same group. Notably, Head, Middle, and Tail institutions in United States show very similar subfield composition to those in the same groups in mainland China. For example, the mean fractions of Head, Middle and Tail subfields in Head institutions are 37.0%, 35.9%, 27.1% in United States and 35.0%, 40.9%, 24.1% in mainland China. This stability across countries indicates that the subfield effect is dominant. A unit's positional profile is determined by which subfields it contains, and this relation is likely to be independent of the institution or country in which those subfields are embedded.

Table 1 lists the top ten Head-enriched subfields shared by the most Head-countries and Head-institutions. The counts are derived from the country × subfield and institution × subfield intersections described above. Neurology & Neurosurgery is the most frequent Head-enriched subfield at both levels (in 19 countries and 31 institutions), followed by Nanoscience & Nanotechnology (in 10 countries and 29 institutions). These are the subfields in which a Head position is most consistently reproduced across countries and institutions, and they represent the research directions that currently sit most consistently at the head of the Top 2% ranking.

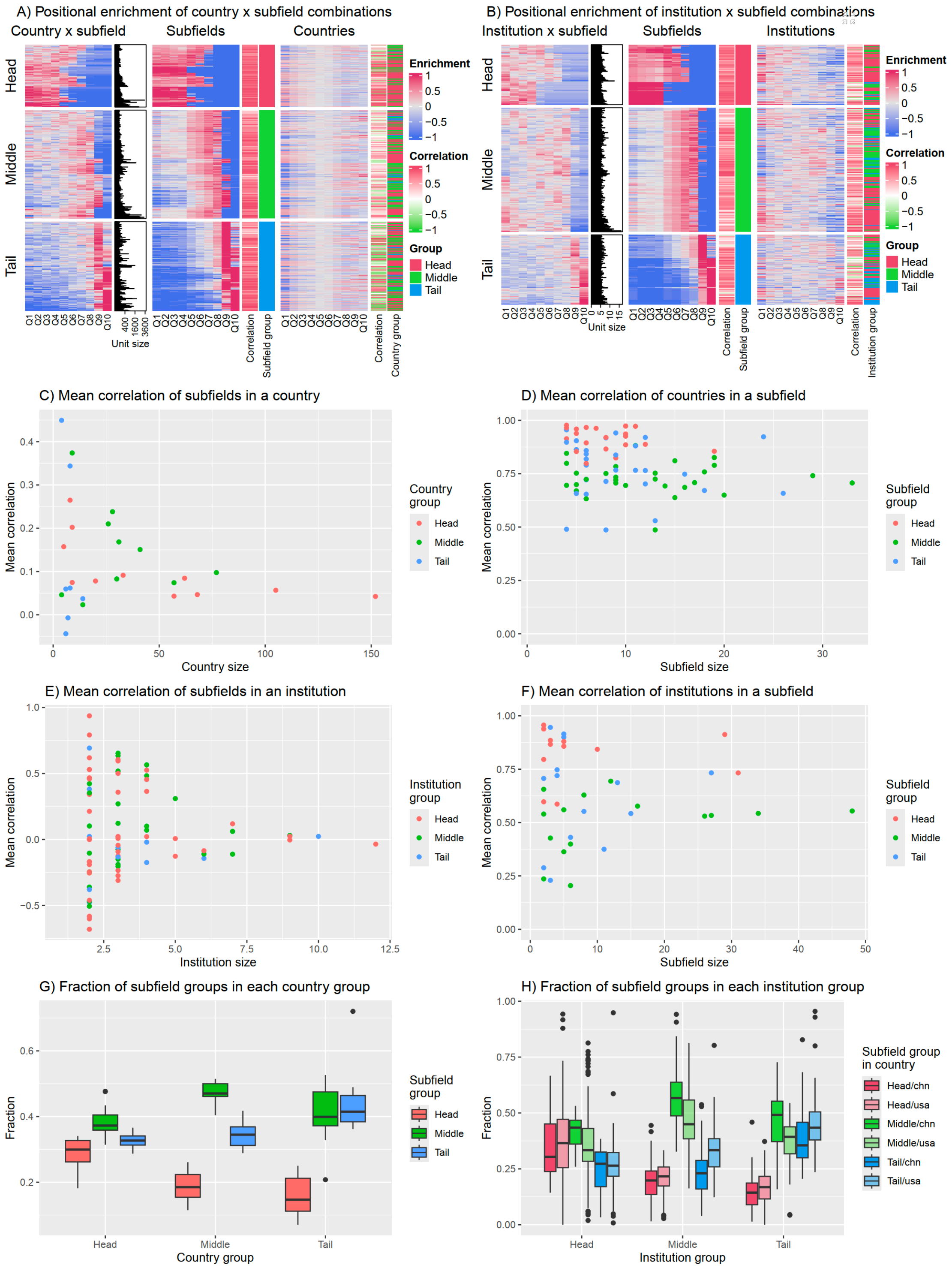

**Figure 4. Positional enrichment of country × subfield and institution × subfield combinations**. (A) Positional enrichment profiles of country × subfield combinations, together with the corresponding country and subfield profiles. (B) Positional enrichment profiles of institution × subfield combinations, together with the corresponding institution and subfield profiles. For both panel A and B, heatmap rows are grouped by subfield clusters. (C) Mean pairwise correlation of subfield enrichment profiles within a country. (D) Mean pairwise correlation of country enrichment profiles within a subfield. (E) Mean

pairwise correlation of subfield enrichment profiles within an institution. (F) Mean pairwise correlation of institution enrichment profiles within a subfield. (G) Fraction of subfields in each enrichment group within each country group. (H) Fraction of subfield groups in institution groups in United States and mainland China.

| Subfield | #Country | #Institution |
|---|---|---|
| Neurology & Neurosurgery | 19 | 31 |
| Endocrinology & Metabolism | 12 | 1 |
| Ecology | 11 | - |
| Nanoscience & Nanotechnology | 10 | 29 |
| Developmental Biology | 10 | 10 |
| Environmental Sciences | 10 | 5 |
| Geochemistry & Geophysics | 10 | 2 |
| Business & Management | 10 | - |
| Education | 9 | 4 |
| Food Science | 8 | 5 |

**Table 1. Top ten Head-enriched subfields in Head-countries and institutions.** Numbers were collected from the two Head row group in Figure 4A and 4B.

## High impact requires approximately ≥ 25 years of academic age

We defined academic age as the number of years since a scientist's first indexed publication and grouped scientists into five-year age bins. We computed the positional enrichment profile for each age unit, with minimal number of 500 scientists per bin, which results in the 70-74 bin being the oldest age group. The complete positional enrichment table can be found in Supplementary Table 7.

Figure 5A shows that young scientists below 25 years of academic age are concentrated at the tail, with younger age, the stronger enrichment towards $Q_{10}$. After 25 years, the profile flips with enrichment shifting rapidly toward $Q_1$ and stronger Head enrichment as academic age increases. The highest Head-enrichment occurs in the age bin of 55-59 (enrichment = 0.889). The pattern is monotonic in both the increasing and decreasing directions, with no reversal at any age, indicating that the age-position relationship forms a consistent gradient across the academic lifespan.

To ask whether the age gradient is mediated by research subfield composition, we computed, for each age bin, the percentage of scientists whose subfield is classified as Head-, Middle-, and Tail-enriched (Figure 5B). Before 25 years, the share of Head-enriched subfields rises steadily with age, while after 25 years, it stabilizes and converges to the global fraction. The Middle and Tail shares show the opposite pattern, both declining with age and stabilizing after 25. This trend reflects the fact that young scientists remain in the same subfield but are still accumulating citations, so they sit at the lower end of their subfield and are less likely to be counted as part of its Head component. As their citation records grow, they move toward the head of their subfield, and the share of Head-enriched subfields in their age bin approaches the subfield's own Head status. The stabilization after 25 years indicates that this process is largely complete.

National and institutional composition follow similar overall pattern (Figure 5C and 5D). At both levels, the fraction of Head units rises with age and exceeds the overall baseline from 25 years onward, indicating Head-enrichment at the country and institution level is an accumulation process. The two levels differ at young ages. At the national level, the Head share is lower than the Middle share before 15-19, indicating that Middle countries contain a higher proportion of young scientists and a lower proportion of senior ones. While at the institution level, the three groups start almost evenly in the 00-04 bin. These patterns differ from the stabilization of subfield composition after 25 years. Subfield is a dimension that scientists enter and remain in, so its composition stabilizes once careers are established, whereas institution and country are dimensions that scientists are located in, and their age profiles reflect the academic-accumulation structure of those units. A comparison of United States (the Head country) and mainland China (the Middle country) illustrates these compositional differences. Figure 5E shows that Chinese scientists are systematically younger, mainly located in the 05–34 age bins, whereas American scientists are more broadly distributed up to the 70-74 bin. Figure 5F shows that the Head-subfield share is almost stable across age bins within each country but systematically lower in mainland China than in United States at every age bin (bins are filtered by minimum of 25 scientists). United States accumulates a higher proportion of senior scientists and a higher proportion of Head subfields, which may push its positional enrichment to the head of the Top 2% list.

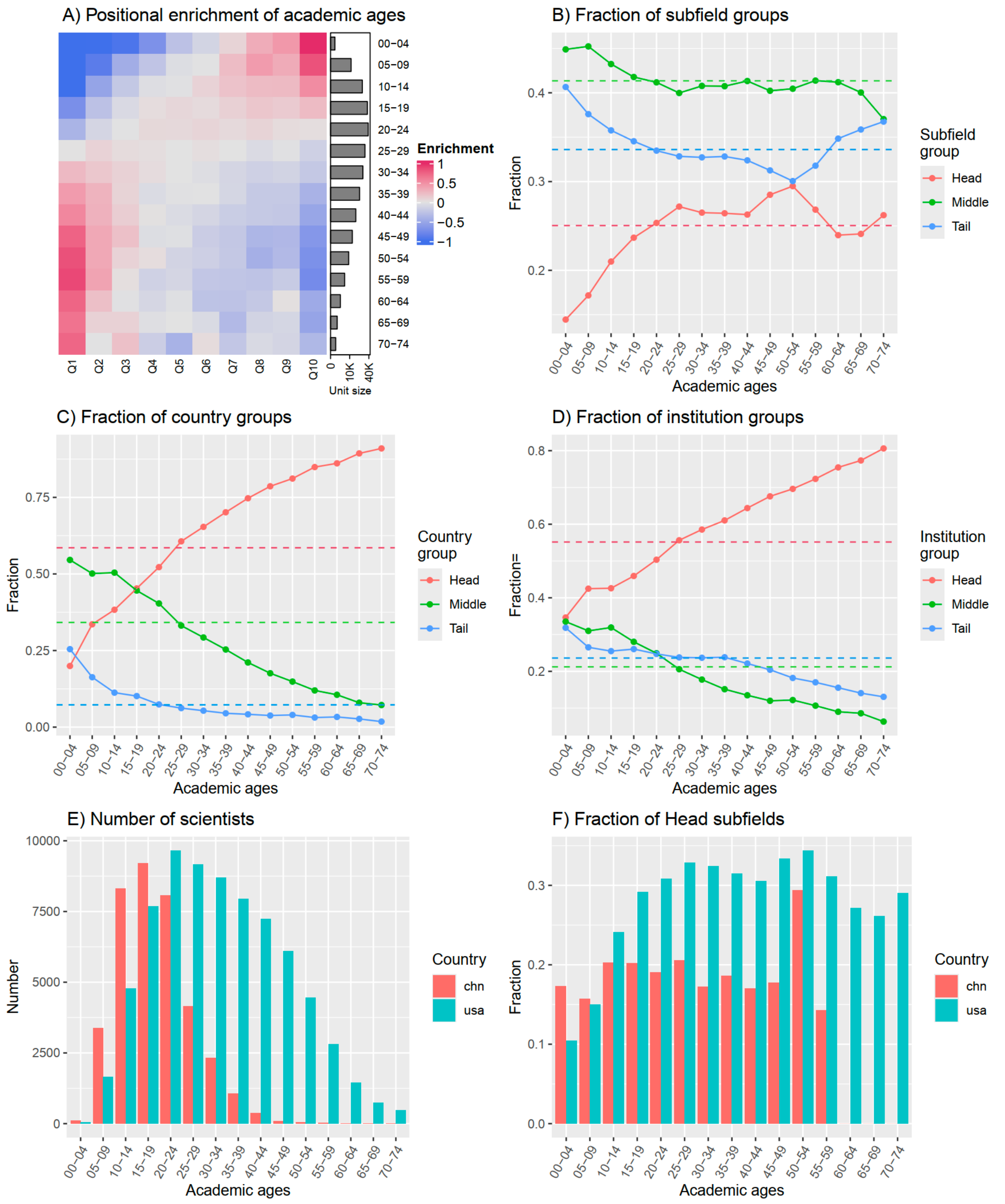

**Figure 5. Positional enrichment of academic ages.** A) Positional enrichment profiles. B) Fraction of subfields in each academic age bin. C) Fraction of country groups in each academic age bin. D) Fraction of institution groups on each academic age bin. In panel B-D, dashed lines are the global fraction baselines of subfields, countries and institutions. E) Number of scientists on age bins in United States and mainland China. F) The Head-subfield fractions on age bins in United States and mainland China.

## Emerging exceptions below 25 years of academic age

The previous section showed that scientists below 25 years of academic age are concentrated at the tail of the Top 2% list. Figure 6 identifies, among these young scientists, the countries, institutions, and institution × subfield combinations whose positional enrichment runs opposite to the overall trend, units that are Head-enriched despite their young academic age. We restricted the analysis to the 00-04, 05-09, 10-14, 15-19, and 20-24 age bins and treated each unit × age-bin combination as a separate unit, requiring a minimum of 50 scientists. We then computed the Spearman correlation between the enrichment profile of each unit × age-bin combination and that of its corresponding age bin, and only kept units with a correlation < -0.3.

Figure 6A shows that only six countries appear as Head-enriched below 25 years. They are small, high-income economies with strong international recruitment and concentrated research investment: the United Arab Emirates, Hong Kong, Macao, Singapore, Switzerland, and Saudi Arabia. At the institutional level (Figure 6B), the same pattern appears with a larger and more diverse set. They are globally highly recognized institutions, e.g., ETH Zürich, MIT, Peking University, the University of Oxford, Stanford University, that show large research volumes, dense international collaboration networks, and well-resourced research environments. These are the factors that allow young scientists to accumulate citations faster than in less research-intensive institutions.

At the institution × subfield level, Figure 6C is dominated by Artificial Intelligence & Image Processing that appears at every age bin from 00-04 to 20-24, carried by a small set of institutions: DeepMind, Google, Meta, NVIDIA, Tsinghua, and Shanghai Artificial Intelligence Laboratory. This observation is consistent with the broader context of AI research in recent years. The field has experienced rapid growth in publication volume and citation accumulation, with short publication-to-citation cycles, dense cross-institutional collaboration, and a substantial industrial research base [19], [20], [21]. These features that are associated with faster citation accumulation among young scientists and that may explain why AI scientists reach the head of the ranking at academic ages far below the 25-year threshold that characterizes the aggregate age trend. The presence of industry laboratories, DeepMind, Google, Meta, NVIDIA, contrasts with the underperformance of industry institutes as a group in Figure 2E, indicating that a citation-dense subfield can override the disadvantage associated with institutional type.

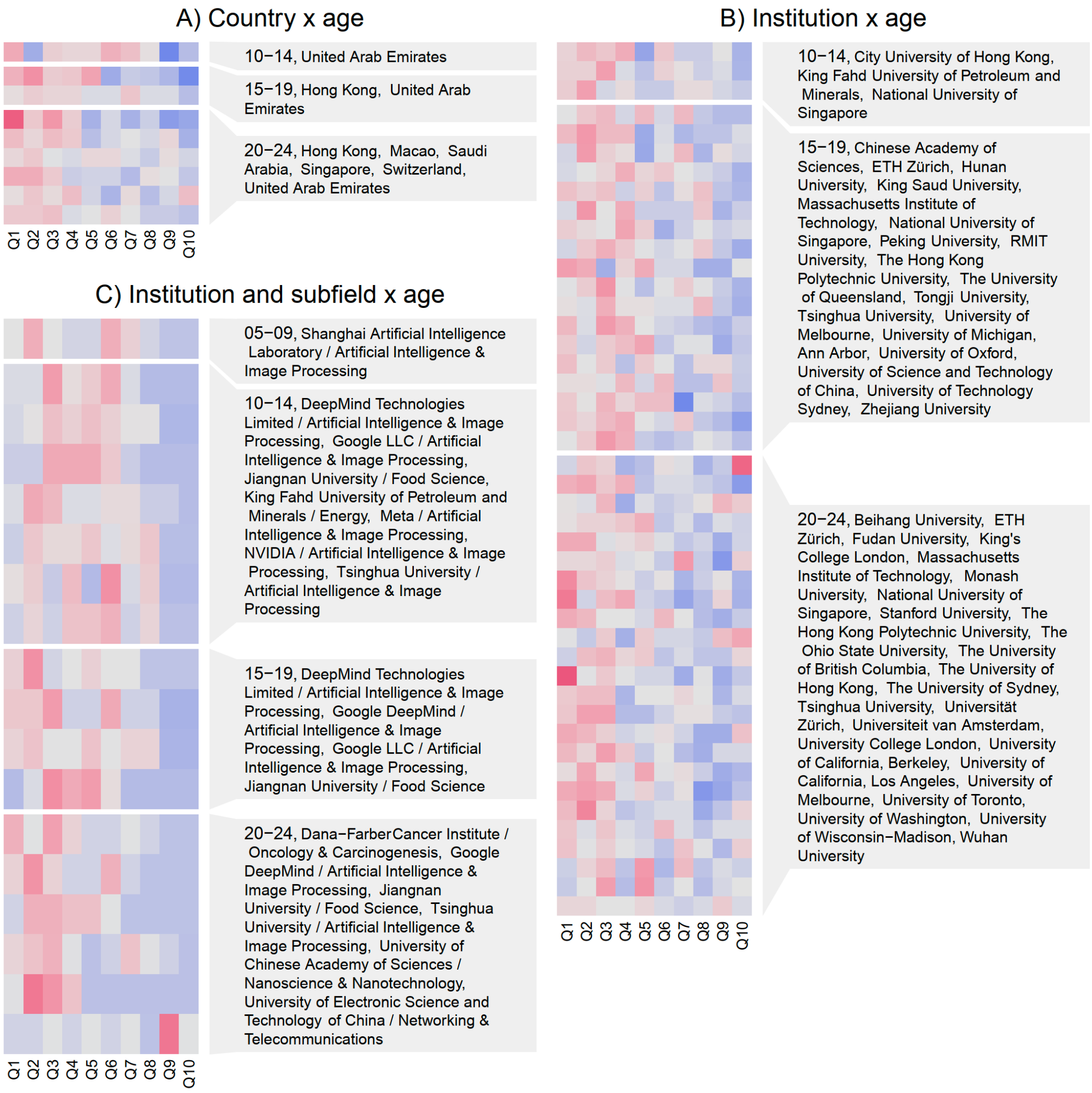

**Figure 6. Head-enriched units in early academic ages.** (A) Country × age-bin combinations. (B) Institution × age-bin combinations. (C) Institution × subfield × age-bin combinations.

# Discussion

In the most typical use of the Top 2% scientists list, a country's or institution's standing is summarized by a single number, which is how many of its scientists appear on the list. That number answers a question of size, but it provides no information about where those scientists sit. A country with many listed scientists may have them concentrated at the very top, spread evenly across the ranking, or pushed toward the bottom, and these three situations are indistinguishable in a raw count. Understanding this internal structure is a precondition for interpreting what the list actually measures and for identifying the factors that separate scientists along the ranking.

We proposed the positional enrichment method to answer the question: how are a unit's scientists distributed along the list? The method has two advantages for analyzing rankings in bibliometrics studies. First, by comparing to a uniform baseline where scientists are randomly placed and are independent of countries or research fields, positional enrichment can reveal where a unit's scientists preferentially sit and allows that preference to be examined at different levels. Second, the enrichment metric is relative and normalized, so it is insensitive to unit size and can be used to compare units of very different scales. In this study, we used positional enrichment to reveal the internal structure of the ranking across multiple levels and dimensions. One caveat should be noted, however. When interpreting the measure, a negative value should not be read as few or no scientists being located in that region. It means that the number of scientists there is lower than expected given the unit's overall size.

The comparison against a random background in bibliometrics studies can be traced back to the mid-1980s. In study [22], the Relative Citation Rate (RCR) was proposed to measure the citation activity of a unit, defined as the ratio of the observed number of citations to the expected number, where the expected value is the average over the background union of all units. As the original publication states: "*Under proper conditions, RCR makes possible direct comparison and even linear ranking of citation impact of publications in different fields.*" Subsequently, numerous studies proposed similar relative metrics [23], [24], [25], [26], [27]. Although not always stated explicitly, comparison against a background implicitly assumes that citations are distributed uniformly across the background. In our previous study [14], we also implemented this idea, *citation enrichment*, to examine the directional citation preference of one country toward another, and we successfully revealed a global two-world structure separating scientifically developed from developing countries, and showed that it is insensitive to their scientific volumes.

Applying positional enrichment across multiple levels, the most significant finding of this study is that structural differences across levels are organized primarily by research subfields. The same subfield occupies similar region of the ranking whether measured globally, within a country, or within an institution, and its profile is highly consistent across these levels. National and institutional differences are therefore a compositional effect. Head countries and Head institutions sit at the head because they contain larger shares of Head subfields, not because their scientists behave differently within a subfield. This has an important implication for evaluation. A unit's position in the list reflects its disciplinary composition, not its research quality across all fields. Citation density differs substantially across research subfields. For example, clinical medicine and biomedical research are far more citation-dense than the humanities and social sciences. Because the Top 2% list is built on citation-based indicators, citation-dense subfields are therefore expected to be over-represented toward the head of the ranking. Cross-disciplinary comparisons of positional profiles should therefore be interpreted with caution.

The dominant role of subfield is associated with a three-group pattern of positional enrichment profiles that is clear and consistent across countries, institutions, and fields. Head-enriched units are over-represented in the leading bins, Middle-enriched units in the middle-to-late bins, and Tail-enriched units in the final bins. At the national level, the three-group classification of countries shows high agreement to our previous study [14], in which countries were clustered by mutual citation preference, that is, for a paper from country *A*, given two comparable papers from countries *B* and *C*, which of the two *A* prefers to cite. Clusters were defined so that countries within a cluster show strong mutual citation preference. Using the group labels defined in [14], the Head group overlaps substantially with West-1 (developed English-speaking and Western European countries), the Middle group with Asia-1 (advanced Asian countries), Europe-2 (Spain and Italy), and Europe-3 (Southern and Eastern European countries), and the Tail group with Asia-2 (Southeast and Southern Asian countries), the Middle East, and Latin America. A natural interpretation is that the positional distribution of a country is shaped by the citation behavior of its peer countries in the same geographical region and in the common subfields they share. Since the ranking is built from citation records, one testable direction for future work is whether countries in the same geographical region are more active in the same citation-dense subfields. If so, the positional grouping would reflect a citation-flow structure in which countries reinforce each other's visibility within shared research areas.

The academic age analysis adds a complement layer, where we found a transition of the positional enrichment profiles around 25 years. This may be explained by the research career development. From graduate student to independent senior researcher typically takes decades or more. This suggests that the age threshold is not purely an effect of citation accumulation but also reflects the career stages built into the research system. Subfield composition stabilizes after 25 years, whereas institutional and national composition continue to shift across the career. Subfield is a dimension that scientists enter and remain in, while institution and country are dimensions in which they are located, and their age profiles reflect the senior/junior composition of the units themselves. The stability of subfield composition and the continued shift of institutional and national composition are therefore not contradictory but correspond to two different mechanisms, which is, positioning along a dimension versus age stratification within units. The comparison between United States and mainland China may explain why

Head countries and Head institutions sit at the head: they contain larger shares of Head subfields, and these subfields tend to have a higher proportion of senior scientists, so that both composition and the seniority structure within subfields contribute to their head position. At the same time, the interaction between age and subfield suggests that the 25-year transition time point may not be uniform among subfields. Different subfields may have different age distributions, and AI shows the clearest case of an earlier transition toward the head.

Several limitations and potential directions for future work remain. First, although the six components of the c-score are standardized globally, the list is assembled from two selection criteria: a global c-score threshold (top 100,000) and a subfield-specific percentile threshold (top 2% within each subfield). This dual selection directly determines how many scientists from each subfield enter the list, but whether it also acts as a co-factor behind the dominance of subfields in positional enrichment requires further investigation. Second, positional enrichment requires a sufficient number of scientists per unit. As the stratification becomes finer, for example, by age bin, by institution within subfield, or by country within subfield, the number of scientists per bin decreases, the variance of the enrichment estimates increases, and the profile becomes less reliable. This is why we require minimum-size thresholds throughout, but the thresholds themselves limit how finely the analysis can be applied. Third, each of the six components of the c-score can itself define a ranking, and positional enrichment could be applied to each component-specific ranking to examine how positional distributions differ across citation indicators. Since the single-author citation component shows a large difference from the other five components [2], it is natural to hypothesize that a ranking based on single-author citations would produce a different positional picture, and to explore the mechanism behind this difference. Forth, the academic-age analysis, and more generally any analysis of how positional profiles changes over time, requires longitudinal data. Our cross-sectional comparison across age bins is based only on the static 2026 list and describes how scientists of different academic ages are distributed, but it cannot establish how positional profiles change over an individual career. Addressing these gaps, particularly through longitudinal data and component-specific rankings, would further clarify how positional structure is produced, sustained and changed over time.

# Conclusion

Using positional enrichment, we conducted an in-depth analysis of the internal structure of the Top 2% ranking across countries, institutions, research fields, and academic ages. The analysis identified research subfield as the organizing dimension of the ranking, and showed that national and institutional differences in positional placement are driven by composition, not by how scientists perform within a subfield. It also revealed a systematic relationship between academic age and positional placement, with a transition around 25 years and an earlier transition in fast-moving subfields. More broadly, positional enrichment provides a general framework for ranking-based analyses in bibliometrics. By shifting attention from how many scientists a unit contributes to where those scientists sit, it makes visible a dimension of ranking structure that count-based summaries cannot capture, and it offers an extensible approach for studying rankings in other evaluation contexts.

# Supporting information

S1 File: Positional enrichment profiles of institutions and those of their countries.

S1 Table: Positional enrichment profiles of countries.

S2 Table: Positional enrichment profiles of institutions.

S3 Table: Positional enrichment profiles of research fields.

S4 Table: Positional enrichment profiles of research subfields.

S5 Table: Positional enrichment profiles of country × subfield combinations.

S6 Table: Positional enrichment profiles of institution × subfield combinations.

S7 Table: Positional enrichment profiles of academic ages.

# Acknowledgments

This research did not receive any specific grant from funding agencies in the public, commercial, or not-for-profit sectors.

# Author contributions

Conceptualization: Zuguang Gu.

Data curation: Yuehai Li, Zuguang Gu.

Formal analysis: Yuehai Li, Zuguang Gu.

Funding acquisition: Zuguang Gu.

Investigation: Yuehai Li, Zuguang Gu.

Methodology: Zuguang Gu.

Project administration: Zuguang Gu.

Resources: Zuguang Gu.

Software: Zuguang Gu.

Supervision: Zuguang Gu.

Validation: Zuguang Gu.

Visualization: Zuguang Gu.

Writing – Zuguang Gu.

Writing – review & editing: Yuehai Li, Zuguang Gu.